\documentstyle[12pt,aps,epsf]{revtex}
\vspace{2cm}
\begin{document}
\title{\ \\ \ \\ \ \\ \ \\ \ \\  \ \\
INTERACTIONS AND LOCALIZATION:\\ TWO INTERACTING PARTICLES APPROACH}
\author{D.L.Shepelyansky $^{*}$}
\address{
Laboratoire de Physique Quantique, Universit\'{e} Paul Sabatier,\\
118, route de Narbonne, 31062 Toulouse, France
}
\maketitle
\thispagestyle{empty}
\bigskip

{\small It is shown that two repulsing / attracting particles in a random
potential can propagate coherently on a distance $l_c$ much larger than 
one-particle localization length $l_1$ without interaction. In dimension $d>2$
this leads to delocalization of pairs formed by two repulsing / attracting
particles. The results of numerical simulations allow to understand some
specific features of this effect.
}
\bigskip

\section{Introduction}

The problem of interacting particles in a random potential
attracts more and more interest during last years (see \cite{Kirk}
and refs. therein). The situation at finite particle density
is rather complicated both for analytical and numerical
analysis and therefore it is desirable to have some 
relatively simple models which could be solved and 
would lead to a better understanding of effects of interaction in the
presence of disorder and localization. 

One of the ways to analyse the effects of interaction is 
to treat them in a self-consistent
way. The simplified model of this approach can be described by
nonlinear Schroedinger equation on a discrete lattice with disorder
\begin{equation}
i {{\partial {\psi}_{n}} \over {\partial {t}}}
=E_{n}{\psi}_{n}
-{\beta}V{\mid{\psi_{n}}\mid}^2 \psi_{n}
 +V ({\psi_{n+1}}+ {\psi_{n-1})} 
\label{nse}
\end{equation}
where $V$ is the intersites hopping matrix element,
$E_{n}$ are randomly distributed in the interval $(-W,W)$
and $\beta$ represents the dimensionless strength of self-consistent
interaction. The model (1) and other models of a similar type
with nonlinearity and disorder had been studied in \cite{nand,arkady}.
It had been found that below  critical strength of nonlinearity
$| \beta | < \beta_c \sim 1$ Anderson localization persists
while above $\beta_c$ the localization is destroyed and 
anomalous diffusive spearing of probability along the 
lattice takes place \cite{nand,arkady}. Such type of transition
is independent of the sign of $\beta$ and therefore it
looks to be quite different from the results \cite{giam,fisher}
according to which repulsive interaction between particles $(\beta >1)$ 
in solid state systems leads only to a stronger localization near 
the ground state.

The main physical reason for delocalization transition in (\ref{nse})
is due to destruction of interference phase by nonlinear 
term representing interactions in some kind of mean field 
approximation. A more realistic way to study the effect
of decoherence due to interaction is to consider only two
interacting particles (TIP) on a disordered lattice. 
For a short range interaction a mean field approach 
leads to a nonlinear equation of type (\ref{nse}). However,
the investigation of the TIP
quantum problem gives somewhat different result \cite{TIP}.
According to \cite{TIP} two repulsing/attracting particles, 
being on a distance of one-particle localization length $l_1$
from each other, propagate together on a much larger length 
\begin{equation}
{l_{c} } \sim {{l_1}^2 M} { U^2\over {32 V^2}} 
\label{tip}
\end{equation}
where $U$ is the strength of on site interaction between two 
particles, $V$ is the one-particle hopping element between
nearest sites which determines the size of one-particle energy
band. The parameter $M$ notes the number of transverse channels
for the case when both particles are moving in a thick wire,
where by itself $l_1 \propto M$. Also in (\ref{tip}) it is
assumed that the intersite constant $a=1$ and the total TIP energy
is somewhere close to the center of the band so that 
wavevector $k_{F} \sim 1/a=1$. The same expression for $l_c$
was also derived in another way by Imry \cite{Imry}
whose approach is based on the Thouless block picture and
scaling of conductance. The reason for the difference between
the model (\ref{nse})and TIP result (\ref{tip}) is not completely clear.
Probably, the mean field approximation is not exact enough.

The derivation of the result (\ref{tip}) is based 
on some assumptions which look quite natural but still at the moment
have not been rigorously justified. The main of them are
the estimate of interaction induced transition matrix elements
which are assumed to have the same order of magnitude and
a partial neglect of correlations between TIP moving
in the same random potential. Therefore it was quite important
to check the existence of the TIP effect (\ref{tip}).
First numerical simulations indicating the existence of enhancement
for TIP localization length had been presented in \cite{TIP}.
Much more advanced numerical investigations for TIP in one-dimensional
Anderson model had been done by Pichard and coworkers \cite{Pichard}
and von Oppen and coworkers \cite{Oppen}. Their results show that
the exponent $\gamma$ of power growth $l_c \propto {l_1}^{\gamma}$ 
is close to the theoretical value $\gamma=2$. 
The value of $\gamma \approx 2$ was also found by the transfer matrix
technique for a bag model \cite{TIP,Pichard} in which particles
do not interect if the distance between them is less than the bag
size  $B > l_1$. A model with strong attraction between 
particles in a well of size $B \ll l_1$ had been first studied 
analytically by Dorokhov \cite{Dorokhov} who had found that
the propagation length of such strongly coupled particles
could be enhanced. By an extrapolation to $B \sim l_1$
he argued that in this case the localization length
is proportional to ${l_1}^2$. The case with a short range
repulsive/attractive interaction is much less evident and
more detailed numerical simulations are still required especially
since the results \cite{Oppen} give first power of $U$ for $l_c$
instead of $U^2$ in (\ref{tip}).

The TIP effect in higher dimensions $d$ have been studied in 
\cite{Imry,DS94,Borg1,Frahm3d} where it has been  shown that in $d > 2$
TIP pairs can be delocalized below Anderson transition
when all one-particle states are exponentially localized.
At the moment only few results have been obtained for a number
of particles larger than two \cite{PIm,SSush} and for a finite density
of particles \cite{Imry,Oppen1} where the existence of the Fermi level
plays an important role.     

\section{TIP localization}

The derivation of expression (\ref{tip}) can be done in the following 
way. It is convinient to rewrite the Sroedinger equation for TIP
in the basis of noninteracting eigenstates. In this basis the diagonal
part is given by the sum of one-particle eigenenergies
$\epsilon_{m_1} + \epsilon_{m_2}$. The transitions between noninteracting
eigenstates are only due to interaction and their matrix elements are 
given by
\begin{equation}
{U_s} = U {\sum_{{n_1}, {{{\tilde {n}}_1},{n_2}, {{\tilde {n}}_2}} }
{\tilde{R}}^{+}_{n_1,{\tilde {n}}_1, m_{1},{{\tilde {m}}_1}} 
{\tilde{R}}^{+}_{n_2,{\tilde {n}}_2, m_{2},{{\tilde {m}}_2}} 
{\tilde{R}}_{n_1,{\tilde {n}}_1, m^{'}_{1},{{\tilde {m'}}_1}} 
{\tilde{R}}_{n_2,{\tilde {n}}_2, m^{'}_{2},{{\tilde {m'}}_2}} 
\delta_{n_1,n_2} \delta_{{\tilde{n}_1},\tilde{n}_2}}
\label{us}
\end{equation}
where indices $n_{1,2}, \tilde n_{1,2}$ mark correspondingly the positions of 
first and second particles along and transverse a strip with
$M$ channels, $m$-s are the indices of eigenstates without interaction
which mark the maximum of a state along the strip. The matrix $R$
gives the transformation between the lattice basis and one-particle
eigenstates so that 
$\tilde{R}_{n,{\tilde{n}},m,{\tilde{m}}} \approx 
\exp(-{\mid {n - m} \mid}
/{l_{1}}-{i}\theta_{n,{\tilde{n}},m,{\tilde{m}}})
/{\sqrt{M l_{1}}}$ where $\theta$ randomly changes with indices.

Due to exponential decay of $R$ only about 
$(Ml_1)^{1/2}$ terms with random signs contribute in sum (\ref{us})
so that the typical value of $U_s$ for $|m_1 - m_2| < l_1$ is
$U_s \approx U/(M l_1)^{3/2}$ \cite{nand,TIP}. For $|m_1 - m_2| > l_1$ the 
matrix elements decay exponentially fast and at first approximation they
can be neglacted. Thefore, the total number of coupled states 
is $b \sim (M l_1)^2$. All these states are inside the energy band
$4V$ so that the density of coupled states is $\rho_c \sim (M l_1)^2/V$.
Using the Fermi golden rule we can now determine the interaction
induced transition rate
\begin{equation}
\Gamma \sim {U_s}^2 \rho_c \sim {{U^2} \over {V M l_1}}
\label{gamma}
\end{equation}
The typical size of such transitions is of order $l_1$ so that
they give the TIP pair diffusion rate along the strip
\begin{equation}
D_p \sim {l_1}^2 \Gamma \sim (U/V)^2 V l_1/M \sim (U/V)^2 D_1
\label{diff}
\end{equation}
where $D_1$ is one-particle diffusion rate on a short time scale.
In all these estimates it was assumed $U < V$ and therefore
we see that the diffusion rate due to interaction is not
enhanced ($D_p  \leq D_1$) that is in agreement with the numerical
results \cite{TIP,Borg1}. The diffusion of TIP pair arises as the
result of interparticle collisions which destroy quantum interference phase
and give coherent TIP propagation.

Knowing the diffusion rate $D_p$ it is possible to find the
localization length $l_c$ for a pair in a way similar to that
one used for dynamical localization in the kicked rotator 
\cite{1981,physd87}. Indeed, due to diffusion the total number of 
excited noninteracting eigenlevels grows as 
$\Delta N \sim \Delta m_1 (M^2 l_1) \delta E/V$
where $M \Delta m_1 \sim M (D_p t)^{1/2}$ gives the number of excited
sites for the first particle and the additional factor $M l_1$
takes into account that the distance between two particles is approximately
$l_1$. Also generally not all coupled 
nointeracting eigenlevels are excited but only a fraction of levels 
in some energy interval $\delta E$. Usually, $\delta E \sim \Gamma$
\cite{Jaq} but similar to the case of photonic localization
\cite{physd87} the actual value of $\delta E$ does not enter in the
final answer for localization length. All $\Delta N$ levels are homogeneously
distributed in the energy interval $\delta E$ and therefore the
level spacing between them is $\Delta \nu \sim \delta E/\Delta N$.
Due to uncertainty relation between frequency and time
after the time $t^*$ defined from the relation $\Delta \nu \sim 1/t^*$
the discrete nature of the lines in the spectrum is resolved 
and the diffusion, which should have a continuous spectrum, is stopped.
This gives the localization time and the localization length
\begin{equation}
t^* \sim M^4 {l_1}^2 D_p/V^2, \;\;\; 
\Delta m_{1,2} \approx \Delta n_{1,2} \approx (D_p t^*)^{1/2} 
\approx l_c \sim M^2 l_1 D_p/V \sim l_1 \Gamma \rho_c
\label{t*}
\end{equation}
in agreement with (\ref{tip}). 
The last relation $l_c/l_1 \sim \Gamma \rho_c$ established in \cite{TIP}
is the same as for photonic localization in a complex molecular
spectrum \cite{physd87} with the only difference that here
the size of transition is not the photon frequency but 
one-particle localization length. This relation shows that 
the length $l_c$ is determined by two-particle spread width
$\Gamma$ which can be extracted from the Breit-Wigner
distribution of TIP eigenstates over eigenbasis
of noninteracting particles (see \cite{Jaq} and section V).

It is interesting to note that the final answer for $l_c/l_1$
looks in such a way as interaction is enhanced by the 
squareroot from the number of components $N_1 \sim l_1$ in one-particle 
eigenfuction ($U_{eff} \sim U \sqrt{l_1}$). 
A similar effect had been intensively studied
for enhancement of weak interactions in nulcei \cite{Sush}. However,
there even being enhanced the effect was small and did not give
large physical changes.

\section{TIP delocalization}

Similar approach based on the uncertainety
relation between frequency and time can be used also in higher dimensions $d$.
For that we should take into account that $U_s \sim U/{N_1}^{3/2}$
with $N_1 \sim {l_1}^d$. Therefore $\Gamma \sim U^2/(V N_1)$
and the TIP diffusion rate $D_p \sim {l_1}^2 \Gamma \sim V (U/V)^2
{l_1}^{2-d}$. It is interesting to note that for $d=2$
the diffusion $D_p$ is independent on $l_1$ while for
$d=3$ it decreases with $l_1$ which in its own turn increases when
approaching the one-particle Anderson transition. Due to
diffusion the number of excited levels grows as
$\Delta N \sim (D_p t)^{d/2} {l_1}^d \delta E/V$.
The level spacing is $\Delta \nu \sim \delta E/\Delta N$ and should
be compared with frequency resolution $1/t$.
For $d=2$ the ratio $1/(\Delta \nu t) \sim (U/V)^2 {l_1}^2$ is
independent on time and as usual in $d=2$ the localization
length is proportional to the exponent of this ratio:
\begin{equation}
\ln (l_c/l_1) \sim (U l_1/V)^2 > 1
\label{2d}
\end{equation}
Since in $d=2$ the localization length $l_1$ grows exponentially with
decrease of disorder ($\ln l_1 \sim (V/W)^2$) the enhancement (\ref{2d})
is enormous.

In $d=3$ the spacing $\Delta \nu$ decreases faster than $1/t$
and therefore the TIP pair will be delocalized if at the moment
$t' \sim {l_1}^6/V$, which is determined by two particle level spacing
for a block of size $l_1$  and during which diffusion is always 
going on, the value of $\Delta \nu$ is less than $ 1/t'$. This gives
the condition of TIP pair delocalization in $d=3$ while one-particle states
are well localized (see also \cite{Imry,Borg1}):
\begin{equation}
(U/V)^2 {l_1}^3 > 1
\label{3d}
\end{equation}

While the results (\ref{2d}),  (\ref{3d}) are qualitatively correct
they however don't take into account the effect of possible
pair size growth with time which  for a first
time was discussed in \cite{Borg1}. Indeed, the above 
derivation of $\Gamma$ and $D_p$ is local
and it assumes that the TIP pair size is always of the order of $l_1$.
This would be correct for a bag model in which particles are
confined in a bag of size $l_1$ with infinite walls. But for our
short range interaction the separation of particles is not
excluded. Indeed, there are always matrix elements $U_{-}$ which give 
an increase of the pair size $n_- = |n_1 - n_2|$.
Due to exponential decrease of the operlapping probability
these transitions decay exponentially with $n_-$ as
$U_- \sim U \exp(- n_-/l_1) /{l_1}^{3d/2}$ that gave the reason
to neglact them in the above consideration. However, the existence
of such transitions should definitely produce a slow diffusive pair size
growth \cite{Borg1} ${n_-}^2/t \sim D_- \sim D_p \exp(-2n_-/l_1)$.
This gives the logarithmic growth $n_- \sim l_1 \ln t / 2$ which
should also change the diffusion rate $D_p$ of pair propagation.
Qualitatively the modification of $D_p \rightarrow \tilde{D}_p$ can be
understood in the following way. Since the pair size becomes
in $\ln t$ larger the probability of particles collisions
which from ergodicity is inversely proportional 
to the pair volume becomes in $(\ln t)^d$ times smaller.
The probability of collisions is proportional to $\Gamma$
so that finally $ \tilde{D}_p \sim D_p/(\ln t)^d$ where we
assumed that still the typical size of transition is $l_1$.
Therefore, the average square of displacement of the center of mass of 
the pair $\sigma_+ = <(n_1 + n_2)^2>/4$ grows in a subdiffusive
way $\sigma_+ \sim D_p t / ( \ln t )^d$. Here the power of 
$\ln t$ is the same as in \cite{Frahm3d} where it was obtained 
on the basis of supersymmetry approach. However, 
the stickings in the regions with $n_- \gg l_1$ can lead to quite large
fluctuations and therefore a more rigorous analysis of
logarithmic corrections is still desirable.

The numerical simulations for TIP in 3d are very heavy and at the moment
there are only numerical results obtained in \cite{Borg1} for the model
of two interacting kicked rotators in effective 2-3 dimensions.
The evolution operator of the model  is
\begin{equation}
\begin{array}{c}
{\hat S_2} = \exp \{ -i [
 H_{0}({\hat n})+H_{0}({\hat n'})+U\delta_{n,n'}] \} \\
\times \exp \{-i [ V(\theta,t) + V(\theta',t) ]\}
\end{array}
\label{twokr}
\end{equation}
with ${\hat n}^{(')}=-i {\partial}/{\partial {\theta^{(')}}}$.
Here $H_{0}({ n})$ is a random function of $n$ in the interval
$[0,2\pi]$ and it describes the unperturbed spectrum of rotational phases.
The perturbation $V$ gives the coupling between the unperturbed levels
and has the form 
$V(\theta,t)= k(1+\epsilon \cos\theta_1 \cos\theta_2 \cos\theta_3) \cos\theta$
with $\theta_{1,2,3}=\omega_{1,2,3} \;t$. For incommensurate frequencies
one can go to the extended phase space by replacing
$H_{0}(n) \rightarrow H_{0}(n)+\omega_1 n_1 +\omega_2 n_2 + \omega_3 n_3$
where new actions $n_{1,2,3}$ are conjugated to phases $\theta_{1,2,3}$.
We used $\omega_1=2\pi\lambda^{-1}$, $\omega_2=2\pi\lambda^{-2}$ with
$\lambda=1.3247...$ the real root of the cubic equation
$x^3-x-1=0$  and 
$\omega_{3} = 2\pi/{\sqrt{2}}$.
Then without interaction the effective dimension for one rotator is 
$d=4$ and at fixed $\epsilon > 0$ the one-particle delocalization takes 
place for $k > k_{cr}$. With switched on interaction $U$ the total 
dimension of the extended phase space is five for two rotators so
that it is possible to say that each rotator is moving in effective dimension
$d_{eff} = 5/2$. According to the above picture of TIP delocalization
the pair in (\ref{twokr}) can be delocalized due to interaction
for $k < k_{cr}$. An example of such delocalization 
for second moments $\sigma_+ = <(n+n')^2 /4>$ and
$\sigma_- = <(n-n')^2>$ is shown in Fig.1.
\begin{figure}
\centerline{\epsfxsize 10cm \epsffile{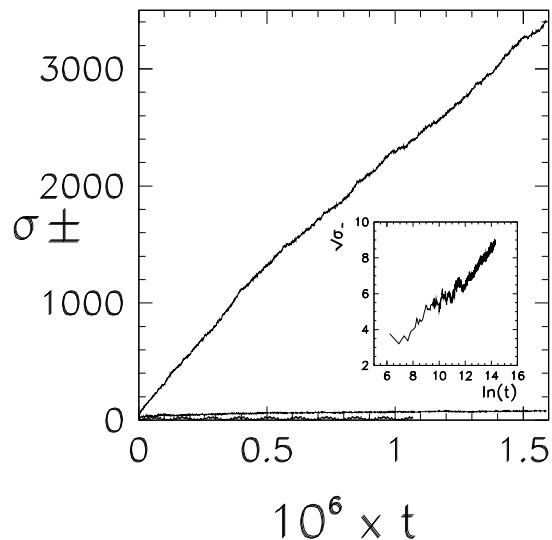}}
\caption{
Dependence of second moments on time
in model (\protect\ref{twokr}) with 
$k=0.7$, $\epsilon=0.9$ $(k_{cr} \approx 1.15)$;
upper curve is $\sigma_+ $ ( $ U=2$ ), middle is $\sigma_-$ ($ U=2 $),
lower is $\sigma_+$ ( $ U=0 $). At $t=0$
both particles are at $n=n'=0$, basis is
$-250 \leq n,n' \leq 250$.  Inset shows the dependence
of ${\sigma_-}^{1/2}$ on $\ln(t)$ (after \protect\cite{Borg1} b).
}
\label{Fig. 1}
\end{figure}
\noindent
In agreement with the above theoretical arguments the pair size
grows logarithmically with time as $n_- \approx \sqrt{\sigma_-} \sim \ln t$.
The behaviour of $\sigma_+$ is consistent with 
$\sigma_+ \propto t/{\ln t}$. In the model (\ref{twokr}) 
the interaction is only
along one direction and therefore the probability of collision
decreases as $1/n_- \sim 1/{\ln t}$ (and not as $\ln ^{-3} t$ 
for real $d=3$) that explains the dependence 
$\sigma_+/t \sim 1/n_- \sim 1/{\ln t}$ (see also \cite{Borg1} b).
The numerical results for two interacting kicked rotators
with incommensurate frequencies clearly demonstrate the
effect of pair delocalization below one-particle delocalization border.

\section{Other models}

Let us now discuss other different models of TIP in a random potential.
A different type of situation corresponds to the case with short
but finite radius of interaction $R < l_1$. In $d$ dimensions the sum 
similar to (\ref{us}) should be taken over $R^d$ nearby sites. 
All these terms have random signs
and therefore the effective value of $U_s$ becomes $R^{d/2}$ times larger
$U_s \sim U R^{d/2}/{N_1}^{3/2}$ where $N_1 \sim {l_1}^d$
is the number of components in one-particle eigenfunction. As the result, the 
enhancement parameter is $\eta \sim [(U/V)^2 R^d] {l_1}^d$ that is
$R^d$ times larger than for on site interaction. As before
the enhancement parameter determines the ratio $l_c/l_1 \sim \eta$ in $d=1$,
$\ln{l_c/l_1} \sim \eta$ in $d=2$ and the TIP pair delocalization 
border $\eta > 1$ in $d=3$. However, it should be taken 
into account that similar to the case $R=1$ the maximal
value of $\eta$ cannot be larger than $N_1$. Indeed, in this case
the interaction is too strong and it starts to deform the
noninteracting density of state. Also it is clear that 
interaction creates some effective wire 
along the diagonal on the lattice of two-particle
index $(n_1,n_2)$ and the number of channels in this wire 
cannot be larger than the number of one-particle components $N_1$.
Therefore, if the parameter $(U/V)^2 R^d$ 
becomes larger than 1 it should be replaced by 1.
For the case of small energies
when $k_F \ll 1/a =1$ the number of independent components
is proportional to $k_F l_1$ and the number of terms in the sum for $U_s$
is of the order of $(k_F R)^d$ so that the enhancement 
is $\eta \sim k_F l_1 (U/V)^2 (k_F R)^d$.  

The above result can be used also for analysis of TIP problem with a long range
interaction. For concreteness let us consider two particles on a distance
$r_{12}$ with Coulomb interaction $\alpha/r_{12}$ in $d=3$ random potential.
Without interaction one-particle eigenfunctions are spreaded over
a size of localization length $l_1$. Interaction two-particle
states mixing appears only in the second order of expansion over
small parameter $l_1/r_{12}$ which corresponds to dipole-dipole
interaction $U_{dd} \sim \alpha {l_1}^2/{r_{12}}^3$. 
Indeed, the first term gives only some locally
homogeneous field which does not destroy localization.
Due to localization the effective radius $R$ of interaction which
determines the number of terms in the sum for $U_s$ is $R \sim l_1$.
Therefore, as above the enhancement factor in $d=3$ is
$\eta \sim [(U_{dd}/V)^2 {l_1}^3] {l_1}^3$ and TIP delocalization
takes place for $\eta > 1$. As before the term in the square brackets
is supposed to be less than 1. It is easy to see that 
TIP can be delocalized even when two particles are very far from each other
$r_{12} \gg l_1$. Due to the homogeneous local field between
particles they will diffusively approach to each other in the case 
of attraction $(\alpha <1)$ or separate up to distance $r_{12}$
where $\eta \sim 1$ in the case of repulsion  $(\alpha > 1)$.

We also can consider other type of TIP problem in which two
particles are moving in parallel strips with independent disorder in each
strip. The number of channels in the strips is $M_1$ and $M_2$
while the localization length for each particle without interaction
is $l_1$ and $l_2$ correspondingly. We will assume that
$M_2 \geq M_1$ and $l_2 \geq l_1$ and that interaction is local
with $U \delta_{n_1,n_2} \delta_{{\tilde{n}_1},\tilde{n}_2}$
where $n_{1,2}$ are the indices along the strips while ${\tilde{n}_{1,2}}$
mark the transverse direction. To estimate the interaction induced
transition matrix elements it is necessary to take into account that
the number of terms contributing to the sum for $U_s$ similar to (\ref{us})
is of the order of $l_1 M_1$ and therefore 
$U_s \sim U/(l_2 M_2 \sqrt{l_1 M_1})$. The density of coupled states
is $\rho_c \sim l_1 l_2 M_1 M_2/V$ and the transition rate 
$\Gamma \sim U^2/(V l_2 M_2)$. The diffusion rate of the first particle
is $D_1 \sim {l_1}^2 \Gamma \sim U^2 {l_1}^2/(V l_2 M_2)$.
In a way similar to the one used above
we obtain the localization length $l_{c1}$ for the first particle 
\begin{equation}
l_{c1}/l_1 \sim \Gamma \rho_c \sim (U/V)^2 l_1 M_1
\label{2strip}
\end{equation}
Surprisingly, $l_{c1}$ does not depend on the characteristics
of the second particle. The localization length for the second
particle is $l_{c2} \approx l_2$ if $l_2 \gg l_{c1}$ and
$l_{c2} \approx l_{c1}$ if $l_2 \ll l_{c1}$.

It is possible to modify slightly the model taking interaction
independent on the transverse direction $U \delta_{n_1,n_2}$ and 
putting $M_1 =1$ (see also \cite{SSush}). Then the number of terms
in the sum of type (\ref{us}) is in $M_2$ times larger. Therefore,
we have the transition rate $\Gamma \sim U^2/(V l_2)$, the diffusion rate
$D_1 \sim U^2 {l_1}^2/(V l_2)$
and $l_{c1}/l_1 \sim (U/V)^2 l_1 M_2$. Such kind of situation in
higher dimension corresponds to the model (\ref{twokr}) where interaction
depends only on one direction and where $l_1 \sim l_2$.
Therefore the TIP diffusion rate in (\ref{twokr})
is $D_1 \sim V (U/V)^2 l_1$ and it grows when approaching
one-particle delocalization border that is in agreement with
numerical data (see \cite{Borg1}b, Fig.13). It is interesting to note
that a similar type of model effectively describes the case
of three interacting particles in a 1d chain where
$M_2 \sim (U/V)^2 l_1$ and three-particle
localization length is $l_{c1}/l_1 \sim (U/V)^4 {l_1}^2 > 1$ \cite{SSush}.

\section{Superimposed band random matrices}

Under some approximations the TIP problem can be reduced to some kind
of band random martix (BRM) model. Indeed, if to write the Hamiltonian
in the noninteracting eigenbasis then it will be represented by
a matrix with a strong diagonal ($\epsilon_{m_1} + \epsilon_{m_2} \sim V$)
and weak ($U_s \sim U/{l_1}^{3/2} \ll V$) 
but broad BRM with approximately $b$ diagonals
where $b \sim {l_1}^2$ is the number of noninteracting eigenstates
coupled by direct transitions and for concreteness we discuss
1d case. Normalizing the nondiagonal elements in a usual 
way (amplitude $\pm 1/\sqrt{2b+1}$) and ordering the levels
which are in the strip of size $\sim l_1$ along levels with $m_1 \approx m_2$
the Hamiltonian matrix will be reduced to a superimposed BRM (SBRM)
with diagonal fluctuations in the interval $\pm W_b$ with
$W_b \sim V \sqrt{l_1}/U$ \cite{TIP}. By transfer matrix technique 
it is easy to investigate the dependence of 
localization length $l_{sb}$ in SBRM
on different parameters. It was shown \cite{TIP} that
for $W_b < \sqrt{b}$ the length scales approximately as
$l_{sb} \approx 0.5 (b/W_b)^2$ while in the perturbative
regime $W_b > \sqrt{b}$ it is $l_{sb} \sim b/{\ln ({W_b}^2 /b)}$.
This result can be understood in a way similar to (\ref{t*}) \cite{TIP}. 
Indeed, the density of coupled states is $\rho_c \sim b/W_b$ and
then the transition rate $\Gamma \sim (1/\sqrt{b})^2 b/W_b \sim 1/W_b$.
As the result, the number of transitions is 
$l_{sb}/b \sim \Gamma \rho_c \sim b/{W_b}^2 > 1$. Taking into account
that for TIP $b \sim {l_1}^2, W_b \sim U \sqrt{l_1}/V$ and
$l_c \sim l_{sb}/l_1$ we can see that the result for SBRM leads
to the same expression (\ref{tip}) for $l_c$.

If the transition rate $\Gamma \sim 1/W_b$ is larger than the
level spacing $1/\rho_c$ then an eigenstate contains many
unperturbed sites with diagonal energies $E_n$ being in the
interval of size $\Gamma$ near the eigenvalue $E_{\lambda}$.
In other words the local density of states 
$\rho_{W}(E-E_{n}) = \sum_{\lambda} \vert \psi_{\lambda}(n) \vert^{2} 
\delta(E-E_{\lambda})$ has the spread width $\Gamma$ 
and is described by the well-known Breit-Wigner distribution \cite{Jaq}:
\begin{equation}
\rho_{BW}(E-E_{n})= 
\frac{\Gamma}{2 \pi((E-E_{n})^{2}+\Gamma^{2}/4)}; \; 
\Gamma = \frac{\pi}{3 W_{b}} 
\label{BW}
\end{equation}
The numerical results \cite{Jaq} confirm that the local density of states
$\rho_{W}$ is well fitted by (\ref{BW}). This result is correct
both for infinite matrix and for matrix of finite size $N < l_{sb}$.
The Breit-Wigner distribution leads to a peaked structure of eigenfunctions
since only levels within $|E_n - E_{\lambda}| < \Gamma$ are populated.
The number of peaks determines the inverse participation ratio (IPR)
$\xi \sim \Gamma \rho$. For delocalized case $N \ll l_{sb}$
the level density is
$\rho \sim N/W_b$ so that $\xi \sim N/{W_b}^2 \ll N$ is much
smaller than the system size. In the localized case $N \gg l_{sb}$ 
the value of $N$ should be replaced by $l_{sb}$ and then the IPR
$\xi \sim l_{sb}/{W_b}^2$ is much less than the localization length
$l_{sb}$ \cite{Jaq}. These results have been also ontained on a more rigorous
basis by supersymmetry approach in \cite{Frahm1,Fyod}.
For original TIP problem this means that the IPR in the
noninteracting eigenbasis $\xi_c \sim (U/V)^4 {l_1}^2 > 1$ is much less than
the number of lattice sites $l_1 l_c$ contributing in an eigenfunction
\cite{Jaq}. Similarly, if one-particle eigenfunction is ergodic
in a $d$-dimensional box of size $L$ then
still the IPR value in noninteracting eigenbasis is much smaller than the total
Hilbert space: $\xi_c \sim \Gamma \rho \sim L^d (U/V)^2 \ll L^{2d}$ \cite{Jaq}.
The existence of Breit - Wigner distribution leads to
a deviation of the number variance $\Sigma^2(E)$ from the random matrix
behavior for energies $E > \Gamma$ where the rigidity of levels
disappeares \cite{Pich1}.

Above we analysed SBRM with bounded fluctuations of matrix elements.
It is interesting to look what will happen if the nondiagonal
matrix elements $H_{n n'}$ have a Cauchy distribution:
$H_{n n'} = \tan (\phi_{n,n'})/\sqrt{2b+1}$ where $\phi_{n,n'}$
is a random phase in the interval $[0,\pi]$ while 
the fluctuations on the diagonal are still bounded $-W_b < H_{n n} < W_b$.
One of the reasons to be interested in such kind of fluctuations
is due to numerical results \cite{Pichard,Flamb} which
have shown that the distribution
of interaction induced matrix elements $U_s$ for TIP problem
on 1d lattice has very long power tails. There are some 
numerical and analytical indications
\cite{Flamb} that $U_s$ can be discribed by a Cauchy
distribution with a typical width $U/{l_1}^2$ and a cutoff
at $U_s > U/l_1$. 

\begin{figure}
\centerline{\epsfxsize 10cm \epsffile{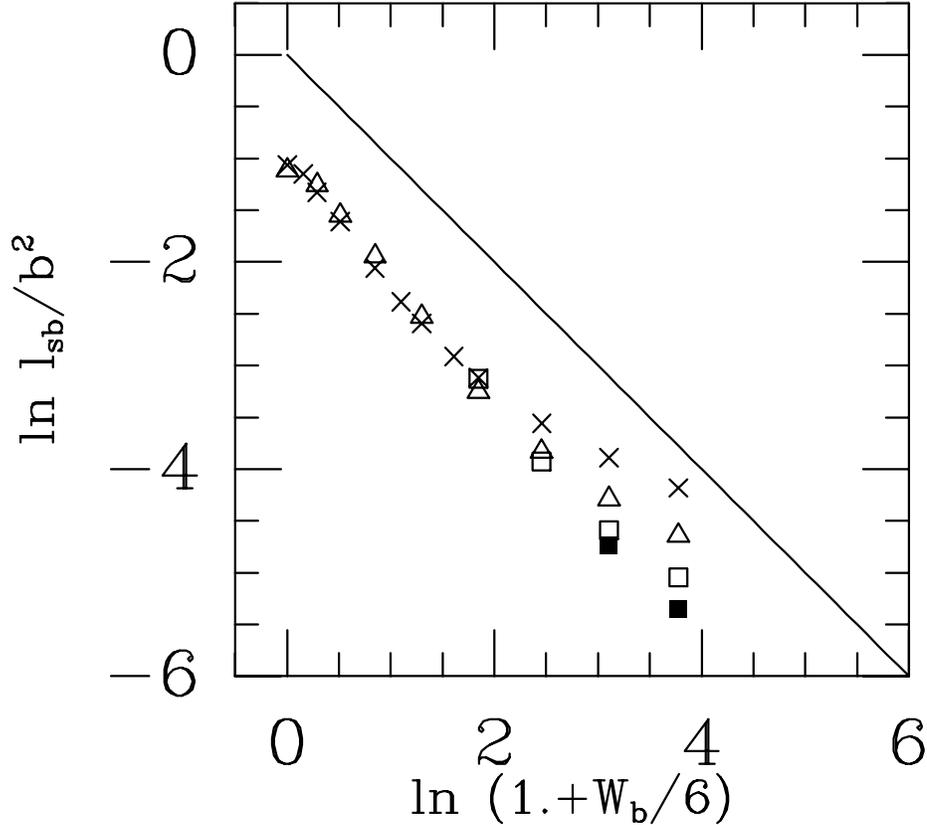}}
\caption{
Dependence of localization length $l_{sb}$ in SBRM with Cauchy
fluctuations on the strength of diagonal fluctuations $W_{b}$
for different band widths with $2b+1 = 41 (X);$ 81 (triangle);
161 (open square); 321 (full square). Full line shows 
the slope -1.
}
\label{Fig. 2}
\end{figure}

\noindent
The transfer matrix numerical investigations for SBRM with
off diagonal Cauchy fluctuations defined above show that
localization length in this case scales as
$l_{sb}/b \approx 3b/W_b >1$ (see Fig.2). The $1/W_b$ behaviour is similar
to the case of Lloyd model however its analytical derivation 
still should be done. If to map this result on the 
TIP case assuming that there $U_s$ has the Cauchy distribution with
width $U/{l_1}^2$ then it would give $l_c/l_1 \sim U l_1$ \cite{Flamb}
in agreement with numerical result \cite{Oppen} for the center of the band.
However, for a serious application to the TIP problem more 
rigorous investigations are required.

\section{Conclusions}

Above we discussed the effect of interaction induced
enhancement of localization length or delocalization mainly for only
two or few particles. In the real physical situation the density
of particles is finite and the situation is much more complicated. 
However, it is possible to think that an effect similar to TIP
effect can take place for quasi-particles. The first estimates
for such a case have been done by Imry \cite{Imry}. They indicate
that in 3d case a mobility edge for  pairs of two quasi-particles
near Fermi energy is lower than one-particle mobility edge.
According to Imry such difference in one- and two-particle edges
can be responsable for anomalous dependence of conductance on
temperature observed in the experiments \cite{Zvi}. However,
more detailed investigations in this direction are required to
clearify the situation. An interesting approximate numerical approach
has been developped \cite{Oppen1} for finite density case 
and applied for 1d chain. There the enhancement of $l_c$
can take place only sufficiently far from Fermi energy in agreement
with \cite{Imry}. However, the 3d case still remains open for
investigations. The TIP effect can be also important for
photo-conductance when an excited pair of quasi-particles
is sufficiently far from Fermi edge and suppression of interaction
due to small phase volume disappeares.

Another interesting question is if the TIP effect can take
place in Luttinger liquid. On the first glance it seems to be 
not the case since the dynamics for Luttinger liquid is 
in some sense completely integrable while for TIP a quite important element
was associated with the ergodic structure of eigenfunctions
and nonintegrability. However, a more detailed analysis is 
required to answer this question.

The author is grateful to the Technion, The Weizmann Institute of Science
and the Godfrey Fund at University of New South Wales for hospitality
during the process of work on the above problem. The useful
discussions with S.Fishman, V.Flambaum, Y.Imry and O.Sushkov are 
also acknowledged.

\vspace{2cm}
\begin{center}
\small{\bf INTERACTIONS ET LOCALISATION : \\
APPROCHE PAR UN MOD\`ELE DE DEUX PARTICULES EN INTERACTION} \bigskip
\end{center}
\bigskip

{\small 
On montre que deux particules avec r\'epulsion ou attraction
dans un potentiel al\'eatoire peuvent se propager de fa\c con coh\'erente
sur une distance beaucoup plus grande que la longueur de localisation
d'une particule sans interaction. En dimension $d > 2$
ceci conduit \`a une d\'elocalisation des paires form\'ees
par deux particules avec r\'epulsion ou attraction.
Les r\'esultats des simulations num\'eriques permettent de comprandre
certaines caract\'eristiques sp\'ecifiques de cet effect.
}
\bigskip


\begin{references}
\bibitem[*]{byline1} Also Budker Institute of Nuclear Physics,
630090 Novosibirsk, Russia
\bibitem{Kirk} D.Belitz and T.R.Kirkpatrik, Rev. Mod. Phys. 
{\bf 66}, 261 (1994).
\bibitem{nand} D.L.Shepelyansky, Phys. Rev. Lett. {\bf 70},
1787 (1993); Physica D {\bf 86}, 45 (1995).
\bibitem{arkady} A.S.Pikovsky and D.L.Shepelyansky, unpublished (1995).
\bibitem{giam} T.Giamarchi and H.J.Schulz,  Phys. Rev. B {\bf 37}, 325 (1988).
\bibitem{fisher} C.A.Doty and D.S.Fisher, Phys. Rev. B {\bf 45}, 2167 (1992).
\bibitem{TIP} D.L.Shepelyansky, Phys. Rev. Lett. {\bf 73}, 2607 (1994).
\bibitem{Imry} Y.Imry,  Europhys. Lett. {\bf 30}, 405 (1995). 
\bibitem{Pichard} K.~Frahm, A.~M\"uller-Groeling, J.-L.~Pichard and D.~Weinmann, 
Europhys.~Lett. ~{\bf 31}, 405 (1995);  D.~Weinmann, A.~M\"uller-Groeling, J.-L.~Pichard and K.~Frahm, 
Phys. Rev. Lett. ~{\bf 75}, 1598 (1995).
\bibitem{Oppen} F.~von~Oppen, T.~Wettig and J.~M\"uller,  Phys.~Rev.~Lett.
~{\bf 76}, 491 (1996).
\bibitem{Dorokhov} O.N.~Dorokhov, Sov. Phys. JETP {\bf 71}(2), 360 (1990).
\bibitem{DS94} D.L.Shepelyansky, unpublished (1994).
\bibitem{Borg1} a) F.~Borgonovi and D.L.~Shepelyansky, Nonlinearity 
{\bf 8}, 877 (1995); b) J. de Phys. I France {\bf 6}, 287 (1996).
\bibitem{Frahm3d} K.~Frahm and A.~M\"uller-Groeling, Phys. Rev. Lett. (to
appear).
\bibitem{PIm} J.-L.Pichard and Y.Imry, unpublished  (1996).
\bibitem{SSush} D.L.Shepelyansky and O.P.Sushkov, cond-mat/9603023 (1996).
\bibitem{Oppen1} F.~von~Oppen and T.~Wettig, Europhys. Lett. 
~{\bf. 32}, 741 (1995).
\bibitem{1981} B.V.Chirikov, F.M.Izrailev, D.L.Shepelyansky, 
Sov. Scient. Rev.C (Gordon \& Bridge) {\bf 2}, 209 (1981);
Physica D {\bf 33}, 77 (1988).
\bibitem{physd87} D.L.Shepelyansky, Physica D, {\bf 28}, 103 (1987).
\bibitem{Jaq} P.~Jacquod and D. L.~Shepelyansky, Phys.~Rev.~Lett. ~{\bf 75}, 
3501 (1995).
\bibitem{Sush} O.P.Sushkov and V.V.Flambaum, Usp. Fiz. Nauk {\bf 136}, 3 
           (1982) [Sov. Phys. Usp. {\bf 25}, 1 (1982)].
\bibitem{Fyod} Y.~V.~Fyodorov and A.~D.~Mirlin, Phys.~Rev. ~{\bf B} (to
appear).
\bibitem{Frahm1} K.~Frahm and A.~M\"uller-Groeling, Europhys.~Lett. ~{\bf 32}, 
385 (1995).
\bibitem{Pich1} D.~Weinmann and J.-L.~Pichard, Preprint, Saclay 1996.
\bibitem{Flamb} V.V.Flambaum, D.L.Shepelyansky and O.P.Sushkov,
unpublished (1995).
\bibitem{Zvi} Z.Ovadyahu, J.Phys. C, {\bf 19}, 5187 (1986).

\end{references}
\end{document}